\documentclass[final,5p,times,twocolumn]{elsarticle}
\usepackage{amssymb}
\biboptions{numbers,sort&compress}

 \usepackage{flushend}
 \usepackage{amsmath}
\usepackage[utf8x]{inputenc}
\usepackage[T1]{fontenc}
\usepackage{amsfonts} 
\usepackage{tikz}
\usepackage{amsfonts} 
\usepackage{mathrsfs}
\usepackage{braket}
\usepackage{graphicx}

\def\beq{\begin{equation}}
\def\eeq{\end{equation}}
\def\bsp{\begin{split}}
\def\esp{\end{split}}
\def\bea{\begin{eqnarray}}
\def\eea{\end{eqnarray}}
\def\ba{\begin{array}}
\def\ea{\end{array}}

\def\lb{\left(}
\def\rb{\right)}

\def\l.{\left.}
\def\r.{\right.}

\def\ra{\rangle}
\def\la{\langle}

\def\bo{\bold{k}}
\def\x{\times}

\journal{Solid State Communications}

\begin{document}

\begin{frontmatter}

\title{XY ring exchange model with frustrated Ising coupling on the triangular lattice}

\author[aff1,aff2]{S. A. Owerre}
\ead{solomon@aims.ac.za}


\address[aff1]{African Institute for Mathematical Sciences -- 6 Melrose Road, Muizenberg, Cape Town 7945, South Africa.}
\address[aff2]{Perimeter Institute for Theoretical Physics -- 31 Caroline St. N., Waterloo, Ontario N2L 2Y5, Canada.}


\begin{abstract}
We investigate the nature of a $Z_2$-invariant XY ring-exchange interaction with a frustrated Ising coupling  on the triangular lattice. In the limit of pure  XY ring-exchange interaction, we show that the classical ground state is degenerate resulting from the $Z_2$-invariance of the Hamiltonian. Quantum fluctuations lift these classical degenerate ground states and produce an unusual state whose excitation spectrum exhibits a gapped maximum quadratic dispersion near $\bo=0$  and  vanishes at the midpoints of each side of the Brillouin zone. This result is in contrast to a gapless  quadratic dispersion near $\bo=0$ in the U(1)-invariant counterpart. We also study the effects of frustration when competing with a classically frustrated Ising interaction.  We provide a glimpse into the possible  quantum phases that could emerge. A comprehensive understanding of this Hamiltonian, however,  cannot be elucidated analytically and requires an explicit numerical simulation.\end{abstract}

\end{frontmatter}

\section{Introduction}
The study of quantum spin ice (QSI)  on three dimensional (3D) pyrochlore lattice has attracted considerable attention \cite{Huang,juan,gin, zhi, you,sun1,  sun4,sun4a, udaa, uda, sun5,sun6,sun7}.  Huang-Chen-Hermele \cite{Huang}  have proposed an alternative Hamiltonian for QSI in 3D pyrochlore lattice,  applicable to certain class of $d$- and $f$-electron systems with dipolar-octupolar Kramers doublets. Using dimensional reduction \cite{sun6, udaa, uda}, Carrasquilla {\it et~al.}, \cite{juan} have recently mapped this model to 2D kagome lattice  with a [111] crystallographic field.  They have identified the interaction that promotes a putative quantum spin liquid (QSL) state and uncovered the low-temperature quantum phase diagram using a non-perturbative, unbiased QMC simulations on the kagome lattice \cite{juan}. In this system, competition between the classical Ising frustration and a $Z_2$-invariant ferromagnetic quantum fluctuation lead to a putative QSL state.  Thus, there is a possibility to search for 2D QSL states within a class of pyrochlore quantum spin ice materials. The distinctive feature of the QSI Hamiltonian is the  presence of $Z_2$ symmetry. We have recently studied  the 2D quantum kagome ice Hamiltonian of Carrasquilla {\it et~al.}, \cite{juan}  on the triangular lattice \cite{sow3}, using spin wave theory. An explicit numerical simulation has not been reported at the moment.   However, spin wave theory still captures the interesting properties of the system because quantum fluctuations are suppressed in this model. In principle, there is a possibility of a ring exchange interaction that exhibits a $Z_2$ symmetry, as in the U(1)-invariant  XY model (hard-core bosons) with ring-exchange interactions \cite{an,an1, long, AW, Ar, N,P,Q, G,F, mic, bber}.  The ring exchange quantum spin Hamiltonian is believed to be very important  in Wigner crystals near the melting density \cite{bb,dj,O}.  They also promote interesting quantum properties with rich quantum phase diagram. 

 In this communication,  we consider the competing interactions between a classically frustrated Ising interaction and a $Z_2$-invariant ring exchange interaction. The Hamiltonian can be written as
\begin{align}
H &= J_z\sum_{\la ij\ra}S_{i}^zS_{j}^z+K\sum_{\left\langle ijkl  \right\rangle} \left(S_{i}^{+}S_{j}^{+}S_{k}^{+}S_{l}^{+} + S_{i}^{-}S_{j}^{-}S_{k}^{-}S_{l}^{-}\right),
\label{model}
\end{align}
where $S_{l}^{\pm}= S_{l}^{x} \pm  i S_{l}^{y}$ are the raising and the lowering spin operators respectively. A special feature of this Hamiltonian is that it exhibits only $Z_2$-symmetry  in the $x$-$y$ plane, {\it i.e.,} $\pi$-rotation about the $z$-axis in spin space, $S_\mu^\pm\to-S_\mu^\pm$, $S_\mu^z\to S_\mu^z$; $\mu=i,j,k,l$. The Hamiltonian (Eq.~\eqref{model})  can be studied in any lattice geometry. However,  for bipartite lattices,  Eq.~\eqref{model} is related to a U(1)-invariant model by a $\pi$-rotation about the $x$-axis on two sublattices, {\it i.e.}, $S_{i,j}^\pm\to S_{i,j}^\mp$; $S_{i,j}^z\to -S_{i,j}^z$.    We restrict our analyses to the triangular lattice. Hence, the summation over the ring exchange term runs over the three possible four-spin plaquette orientations on a triangular lattice; see Fig.~\eqref{fig3.1}. We will investigate the distinctive features  of the pure $Z_2$-invariant XY ring-exchange Hamiltonian ($J_z=0$) and its  effects when competing with a classical Ising frustration ($J_z<K$, with $K<0$).  The study of this Hamiltonian is partially motivated by the  quantum phases uncovered in  2D  QSI Hamiltonian \cite{juan} and the recent study of the 2D  QSI Hamiltonian on the triangular lattice \cite{sow3}.


\section{Pure-K model}
  In order to get an insight into the effects of $Z_2$ symmetry of Eq.~\eqref{model}, we consider the pure-$K$ model in Eq.~\eqref{model}, which corresponds to $J_z=0$. In this limit, the resulting Hamiltonian has a related U(1)-invariant counterpart \cite{an,N}.   The important feature of the U(1)-invariant pure-$K$ model  is that  the energy spectra has  a gapless quadratic excitation near $\bo=0$ \cite{N, an,an1}. However, the present model is  devoid of continuous symmetries. The behaviour of the excitation spectra is not known in literatures.  It is interesting to investigate how the excitations behave in the long wavelength limit.  The Hamiltonian, Eq.~\eqref{model}, in this limit ($J_z=0$) can be written explicitly as   
\begin{eqnarray}\label{eqn3.2}
\begin{split}
H_{J_z=0} &=2K\sum_{\left\langle ijkl  \right\rangle}  \left(S_{i}^{x}S_{j}^{x}S_{k}^{x}S_{l}^{x} + S_{i}^{y}S_{j}^{y}S_{k}^{y}S_{l}^{y} \label{Kterm}\right. \\
 & \left. -  S_{i}^{x}S_{j}^{y}S_{k}^{y}S_{l}^{x} -S_{i}^{y}S_{j}^{x}S_{k}^{x}S_{l}^{y}-S_{i}^{y}S_{j}^{y}S_{k}^{x}S_{l}^{x} \right. \\
&  \left. -S_{i}^{x}S_{j}^{x}S_{k}^{y}S_{l}^{y} -S_{i}^{x}S_{j}^{y}S_{k}^{x}S_{l}^{y}-S_{i}^{y}S_{j}^{x}S_{k}^{y}S_{l}^{x}\right).
  \end{split}
\end{eqnarray}

\begin{figure}
\centering
\begin{tikzpicture}
\draw[ultra thick,blue] (9,0)--(10,0);
\draw[dashed,blue] (10.8,3.95) circle (.35cm);
\draw(10.8,4)node[]{$K$};
\draw[dashed,blue] (7.5,3.5) circle (.35cm);
\draw(7.5,3.5)node[]{$K$};
\draw[dashed,blue] (9.3,0.4) circle (.35cm);
\draw(9.4,.33)node[]{$K$};
\draw[solid,blue] (9,0)--(9.5,0.866);
\draw[ultra thick,blue] (9,0)--(8.5,0.866);
\draw[solid,blue] (10,0)--(11,0);
\draw[solid,blue] (10,0)--(10.5,0.866);
\draw[ultra thick,blue] (10,0)--(9.5,0.866);
\draw[solid,blue] (11,0)--(12,0);
\draw[solid,blue] (11,0)--(11.5,0.866);
\draw[solid,blue] (11,0)--(10.5,0.866);
\draw[solid,blue] (12,0)--(13,0);
\draw[solid,blue] (12,0)--(12.5,0.866);
\draw[solid,blue] (12,0)--(11.5,0.866);
\draw[solid,blue] (13,0)--(14,0);
\draw[solid,blue] (13,0)--(13.5,0.866);
\draw[solid,blue] (13,0)--(12.5,0.866);
\draw[solid,blue] (14,0)--(13.5,0.866);
\draw[ultra thick,blue] (8.5,0.866)--(9.5, 0.866);
\draw[solid,blue] (8.5,0.866)--(9,1.732);
\draw[solid,blue] (8.5,0.866)--(8,1.732);
\draw[solid,blue] (9.5,0.866)--(10.5, 0.866);
\draw[solid,blue] (9.5,0.866)--(10,1.732);
\draw(11,0.7)node[]{$\bold{e}_\alpha$};
\draw(10.1,1.2)node[]{$\bold{e}_\beta$};
\draw[solid,blue] (9.5,0.866)--(9,1.732);
\draw[->,>=stealth,thick, black] (10.5,0.866)--(11.5, 0.866);
\draw[solid,blue] (10.5,0.866)--(11,1.732);
\draw[->,>=stealth,thick, black] (10.5,0.866)--(10,1.732);
\draw[solid,blue] (11.5,0.866)--(12.5, 0.866);
\draw[solid,blue] (11.5,0.866)--(12,1.732);
\draw[solid,blue] (11.5,0.866)--(11,1.732);
\draw[solid,blue] (12.5,0.866)--(13.5, 0.866);
\draw[solid,blue] (12.5,0.866)--(13,1.732);
\draw[solid,blue] (12.5,0.866)--(12,1.732);
\draw[solid,blue] (13.5,0.866)--(13,1.732);
\draw[solid,blue] (8,1.732)--(9,1.732);
\draw[solid,blue] (8,1.732)--(8.5,2.598);
\draw[solid,blue] (8,1.732)--(7.5,2.598);
\draw[solid,blue] (9,1.732)--(10,1.732);
\draw[solid,blue] (9,1.732)--(9.5,2.598);
\draw[solid,blue] (9,1.732)--(8.5,2.598);
\draw[solid,blue] (10,1.732)--(11,1.732);
\draw[solid,blue] (10,1.732)--(10.5,2.598);
\draw[solid,blue] (10,1.732)--(9.5,2.598);
\draw[solid,blue] (11,1.732)--(12,1.732);
\draw[solid,blue] (11,1.732)--(11.5,2.598);
\draw[solid,blue] (11,1.732)--(10.5,2.598);
\draw[solid,blue] (12,1.732)--(13,1.732);
\draw[solid,blue] (12,1.732)--(12.5,2.598);
\draw[solid,blue] (12,1.732)--(11.5,2.598);
\draw[solid,blue] (13,1.732)--(12.5,2.598);
\draw[solid,blue] (7.5,2.598)--(8.5,2.598);
\draw[ultra thick,blue ] (7.5,2.598)--(8,3.464);
\draw[ultra thick,blue] (7.5,2.598)--(7,3.464);
\draw[solid,blue] (8.5,2.598)--(9.5,2.598);
\draw[solid,blue] (8.5,2.598)--(9,3.464);
\draw[solid,blue] (8.5,2.598)--(8,3.464);
\draw[solid,blue] (9.5,2.598)--(10.5,2.598);
\draw[solid,blue] (9.5,2.598)--(10,3.464);
\draw[solid,blue] (9.5,2.598)--(9,3.464);
\draw[solid,blue] (10.5,2.598)--(11.5,2.598);
\draw[solid,blue] (10.5,2.598)--(11,3.464);
\draw[solid,blue] (10.5,2.598)--(10,3.464);
\draw[solid,blue] (11.5,2.598)--(12.5,2.598);
\draw[solid,blue] (11.5,2.598)--(12,3.464);
\draw[solid,blue] (11.5,2.598)--(11,3.464);
\draw[solid,blue] (12.5,2.598)--(12,3.464);
\draw[solid,blue] (7,3.464)--(8,3.464);
\draw[ultra thick,blue] (7,3.464)--(7.5,4.330);
\draw[solid,blue] (7,3.464)--(6.5,4.330);
\draw[solid,blue] (8,3.464)--(9,3.464);
\draw[solid,blue] (8,3.464)--(8.5,4.330);
\draw[ultra thick,blue] (8,3.464)--(7.5,4.330);
\draw[solid,blue] (9,3.464)--(10,3.464);
\draw[solid,blue] (9,3.464)--(9.5,4.330);
\draw[solid,blue] (9,3.464)--(8.5,4.330);
\draw[ultra thick,blue] (10,3.464)--(11,3.464);
\draw[ultra thick,blue] (10,3.464)--(10.5,4.330);
\draw[solid,blue] (10,3.464)--(9.5,4.330);
\draw[solid,blue] (11,3.464)--(12,3.464);
\draw[ultra thick,blue] (11,3.464)--(11.5,4.330);
\draw[solid,blue] (11,3.464)--(10.5,4.330);
\draw[solid,blue] (12,3.464)--(11.5,4.330);
\draw[solid,blue] (6.5,4.330)--(7.5,4.330);
\draw[solid,blue] (7.5,4.330)--(8.5,4.330);
\draw[solid,blue] (8.5,4.330)--(9.5,4.330);
\draw[solid,blue] (9.5,4.330)--(10.5,4.330);
\draw[ultra thick,blue] (10.5,4.330)--(11.5,4.330);
\draw[fill= blue] (9,0) circle (1mm);
\draw[fill= blue] (10,0) circle (1mm);
\draw[fill= blue] (11,0) circle (1mm);
\draw[fill= blue] (12,0) circle (1mm);
\draw[fill= blue] (13,0) circle (1mm);
\draw[fill= blue] (14,0) circle (1mm);
\draw[fill= blue] (8,1.732) circle (1mm);
\draw[fill= blue] (9,1.732) circle (1mm);
\draw[fill= blue] (10,1.732) circle (1mm);
\draw[fill= blue] (11,1.732) circle (1mm);
\draw[fill= blue] (12,1.732) circle (1mm);
\draw[fill= blue] (13,1.732) circle (1mm);
\draw[fill= blue] (7.5,2.598) circle (1mm);
\draw[fill= blue] (8.5,2.598) circle (1mm);
\draw[fill= blue] (9.5,2.598) circle (1mm);
\draw[fill= blue] (10.5,2.598) circle (1mm);
\draw[fill= blue] (11.5,2.598) circle (1mm);
\draw[fill= blue] (12.5,2.598) circle (1mm);
\draw[fill= blue] (7,3.464) circle (1mm);
\draw[fill= blue] (8,3.464) circle (1mm);
\draw[fill= blue] (9,3.464) circle (1mm);
\draw[fill= blue] (10,3.464) circle (1mm);
\draw[fill= blue] (11,3.464) circle (1mm);
\draw[fill= blue] (12,3.464) circle (1mm);
\draw[fill= blue] (6.5,4.330) circle (1mm);
\draw[fill= blue] (7.5,4.330) circle (1mm);
\draw[fill= blue] (8.5,4.330) circle (1mm);
\draw[fill= blue] (9.5,4.330) circle (1mm);
\draw[fill= blue] (10.5,4.330) circle (1mm);
\draw[fill= blue] (11.5,4.330) circle (1mm);
\draw[fill= blue] (8.5,0.866) circle (1mm);
\draw[fill= blue] (9.5,0.866) circle (1mm);
\draw[fill= blue] (10.5,0.866) circle (1mm);
\draw[fill= blue] (11.5,0.866) circle (1mm);
\draw[fill= blue] (12.5,0.866) circle (1mm);
\draw[fill= blue] (13.5,0.866) circle (1mm);
\end{tikzpicture}
\caption{Color online.  Triangular lattice with three-plaquette orientations (thick lines). The ring exchange interaction acts on the four sites within each plaquette. $\bold{e}_\alpha$ and $\bold{e}_\beta$ are the primitive vectors on the triangular lattice.}
\label{fig3.1}
\end{figure}
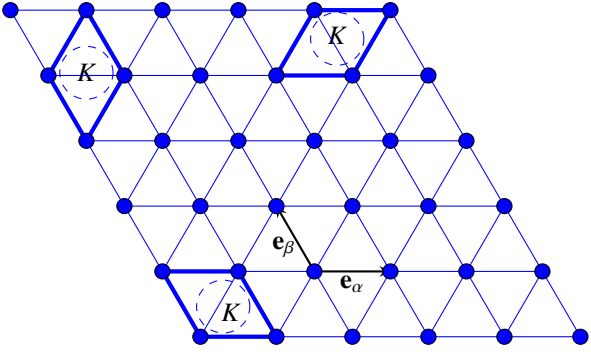

Classically, the ground state of Eq.~\eqref{Kterm} is highly degenerate resulting from the $Z_2$ symmetry of the Hamiltonian. The ground state corresponds to all possible spin configurations  along the basis $\bold{e}_\alpha$ and $\bold{e}_\beta$ (see Fig.~\eqref{fig3.1}), and it is independent of the sign of $K$.
 There are several ways to investigate how quantum fluctuations select a particular classical ground state in this system.  In the U(1)-invariant model, this can be done  by integrating out the  phase fluctuations about the classical ground state in the path integral for the partition function \cite{N}.  This method, however, is effectively the same as performing spin wave theory about any classical ground state of Eq.~\eqref{Kterm}. The collective excitation spectrum is the same in both methods.   
 
 Since we have only $x$-$y$ coupling in Eq.~\eqref{Kterm}, one can show that  the excitation spectrum about any classical ground state is the same, provided one rotates the axes properly.  In the present model  there is no conserved quantity,  it is expedient to use a direct spin wave theory via the Holstein Primakoff transform \cite{J}.  We choose the easy-axis ferromagnetic state, and implement the  linearized Holstein Primakoff transformation, \cite{A,J}
   
 \begin{figure}[ht]
\centering
\includegraphics[width=3in]{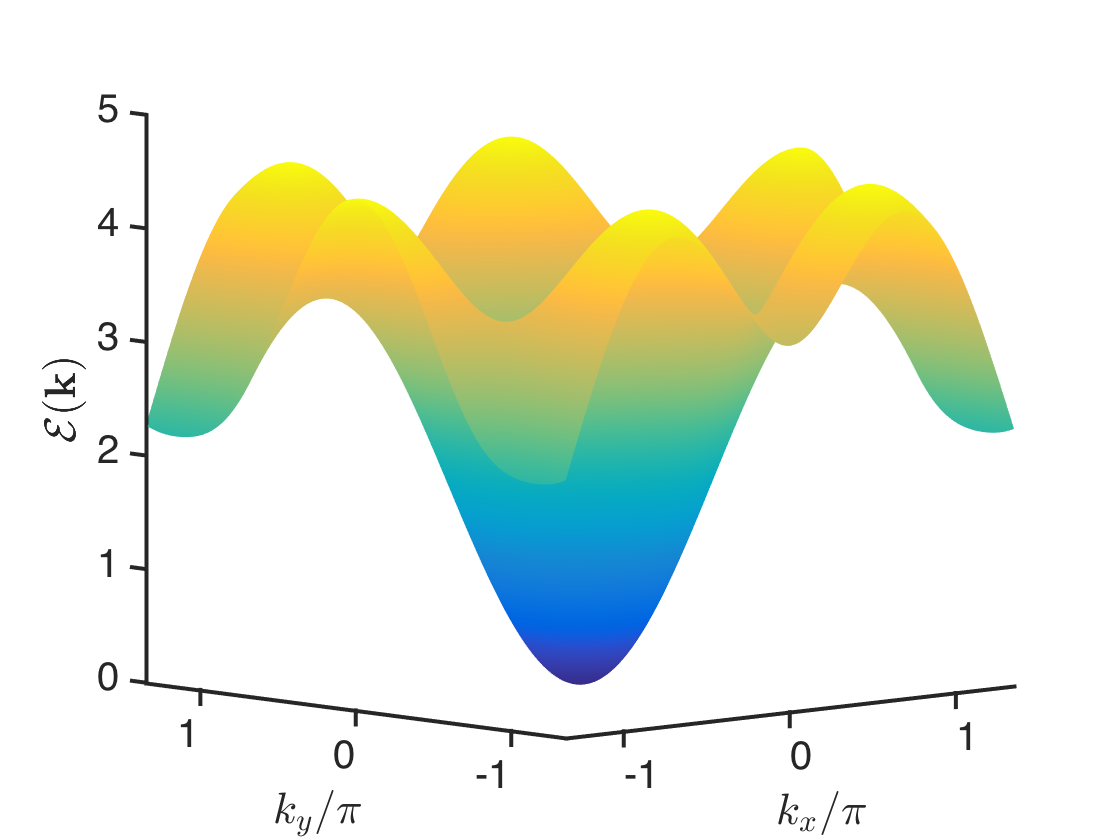}
\caption{Color online.  The spin wave excitation spectrum of the  U(1)-invariant pure-$K$ model. The coupling is set to unity.}
\label{u1}
\end{figure}
   \begin{figure}[ht]
\centering
\includegraphics[width=3in]{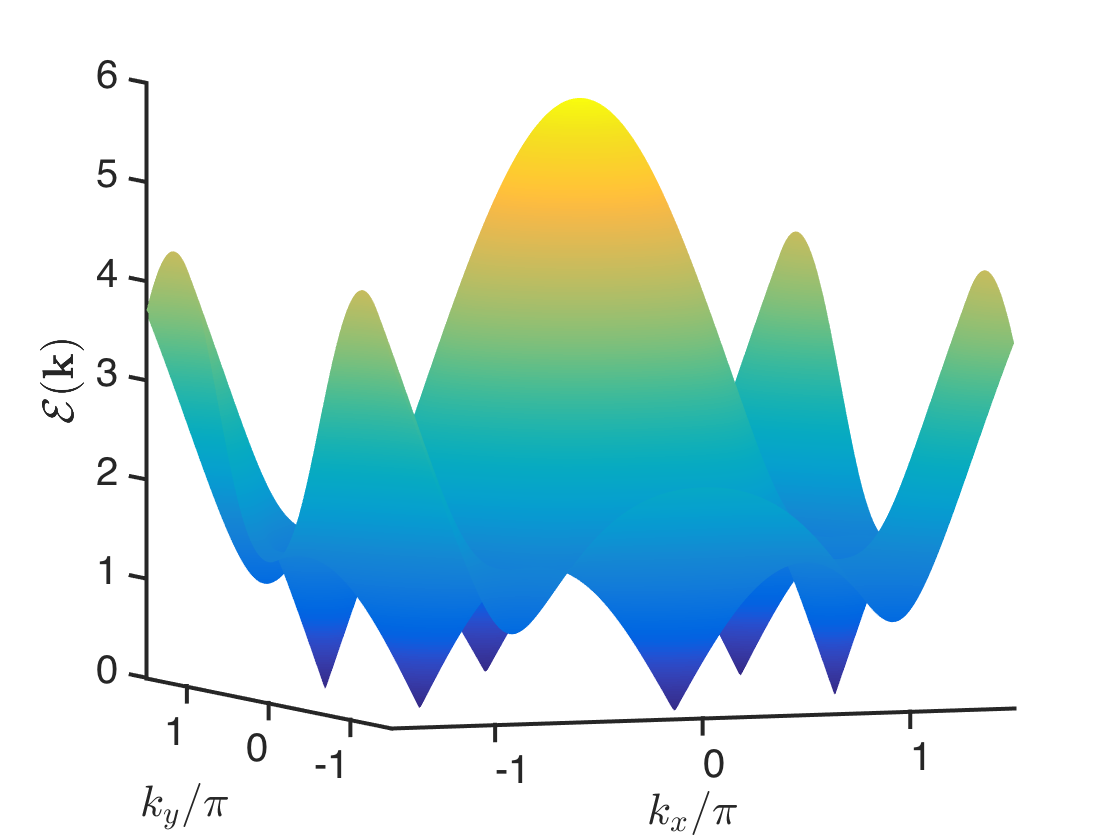}
\caption{Color online.  The energy dispersion of the  $Z_2$-invariant pure-$K$ model. The coupling is set to unity. Fig.~\eqref{zone1} shows the points where the energy vanishes.}
\label{zone}
\end{figure}

 \begin{align}
 &S_{j}^{x}= S-b_{j}^\dagger b_{j },\quad
 S_{j}^{y}= i\sqrt{\frac{S}{2}}\lb b_{j}^\dagger -b_{j}\rb.
 \label{HPT}
 \end{align}
 
  Next, we restrict the spins to $S=1/2$,  substitute Eq.~\eqref{HPT} into Eq.~\eqref{Kterm} and Fourier transform over the three plaquettes.  The resulting bosonic Hamiltonian is very lengthy to write here. It can be diagonalized by the Bogoliubov transformation,
\begin{align}
b_{\bo}=u_{\bo}\gamma_{\bo}-v_{\bo}\gamma_{-\bo}^\dagger,
\end{align}
where $u_{\bo}^2-v_{\bo}^2=1$, one finds that the resulting Hamiltonian is diagonalized by
\begin{align}
&u_{\bo}^2=\frac{1}{2}\lb \frac{A_{\bo}}{E_{\bo}}+1\rb; \quad v_{\bo}^2=\frac{1}{2}\lb \frac{A_{\bo}}{E_{\bo}}-1\rb,\end{align}
with $E_\bo=\sqrt{A_\bo^2-B_\bo^2}$.  

The diagonal Hamiltonian yields
\bea
H_{J_z=0}=\sum_{\bo}E_{\bo}\lb \gamma_{\bo}^\dagger \gamma_{\bo}+\gamma_{-\bo}^\dagger \gamma_{-\bo}\rb.
\eea

The excitation of the quasiparticles is given by

\bea
\mathcal{E}(\bo)=2E_\bo=2\sqrt{A_\bo^2-B_\bo^2},
\eea
\begin{align}\label{eqn3.20}
&{A}_{\bold{k}}= \frac{3K}{2}-B_{\bo},\quad 
{B}_{\bold{k}}=-\frac{K}{2}\lambda _{\bold{k}}- \frac{K}{8}\left(\lambda _{\bold{k}}+\bar{\lambda}_{\bold{k}} \right),
 \end{align}
 and
 \begin{align}\label{eqn3.17}
 &\lambda_{\bold{k}} = \cos k_\alpha+ \cos k_\beta +\cos(k_\alpha +k_\beta);\\&
\bar{\lambda}_{\bold{k}}= \cos(k_\alpha -k_\beta)+\cos(2k_\alpha +k_\beta)+\cos(k_\alpha +2k_\beta).
 \end{align}
where $k_\alpha=\bo\cdot\bold{e}_\alpha$ and $k_\beta=\bo\cdot\bold{e}_\beta$.

\begin{figure}[ht]
\centering
\includegraphics[width=2.5in]{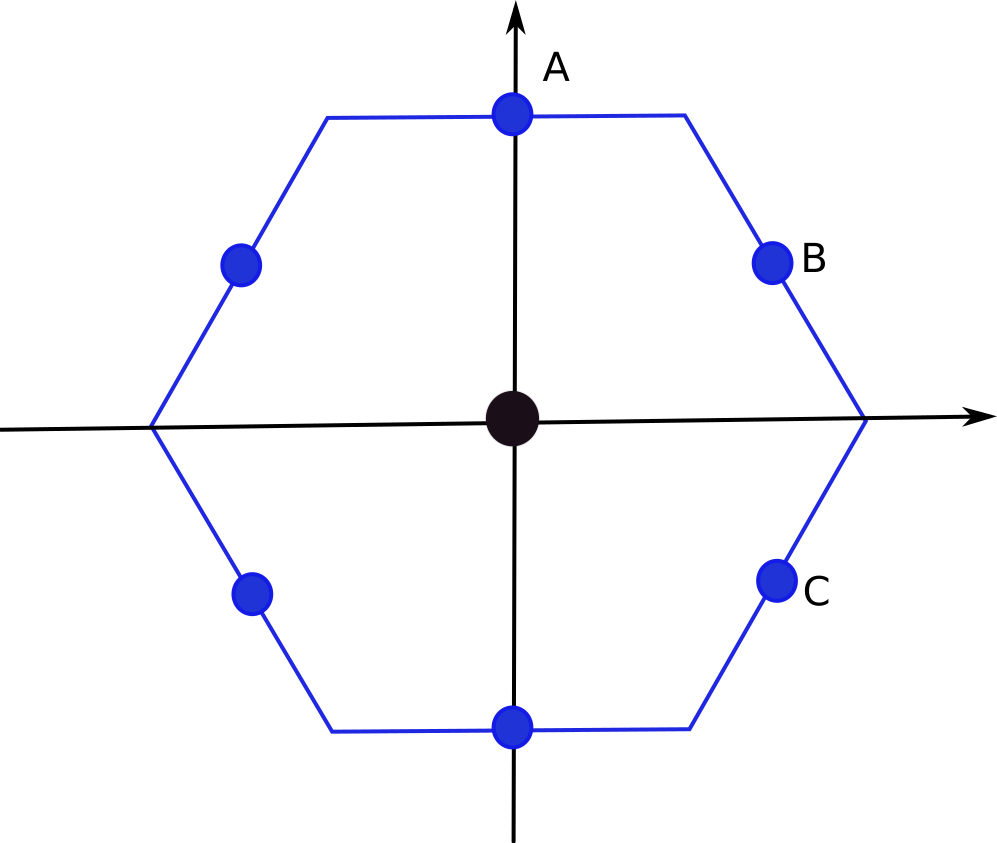}
\caption{Color online.  The Brillouin zone of the triangular lattice. The points $A$, $B$, and $C$ are the midpoints of the adjacent sides of the Brillouin zone, where the energy vanishes.}
\label{zone1}
\end{figure}

In Fig.~\eqref{u1} we have shown the spin wave spectrum of the U(1)-invariant model obtained in Ref.~\cite{N} using a different approach. In this case, there is one quadratic gapless mode at $\bo=0$. In contrast, Fig.~\eqref{zone} shows the spin wave spectrum for the present model. We observe instabilities of the spin wave at the midpoints of the adjacent sides of the Brillouin zone, that is  at the points $A$, $B$, and $C$ in Fig.~\eqref{zone1}.  At the center of the Brillouin zone the spectrum exhibits a gapped maximum dispersion, which in the long wavelength limit behaves as
\begin{align}
\mathcal{E}(\bo) = a-b|\bo|^2,
\label{dis}
\end{align}
\begin{figure}[ht]
\centering
\includegraphics[width=3in]{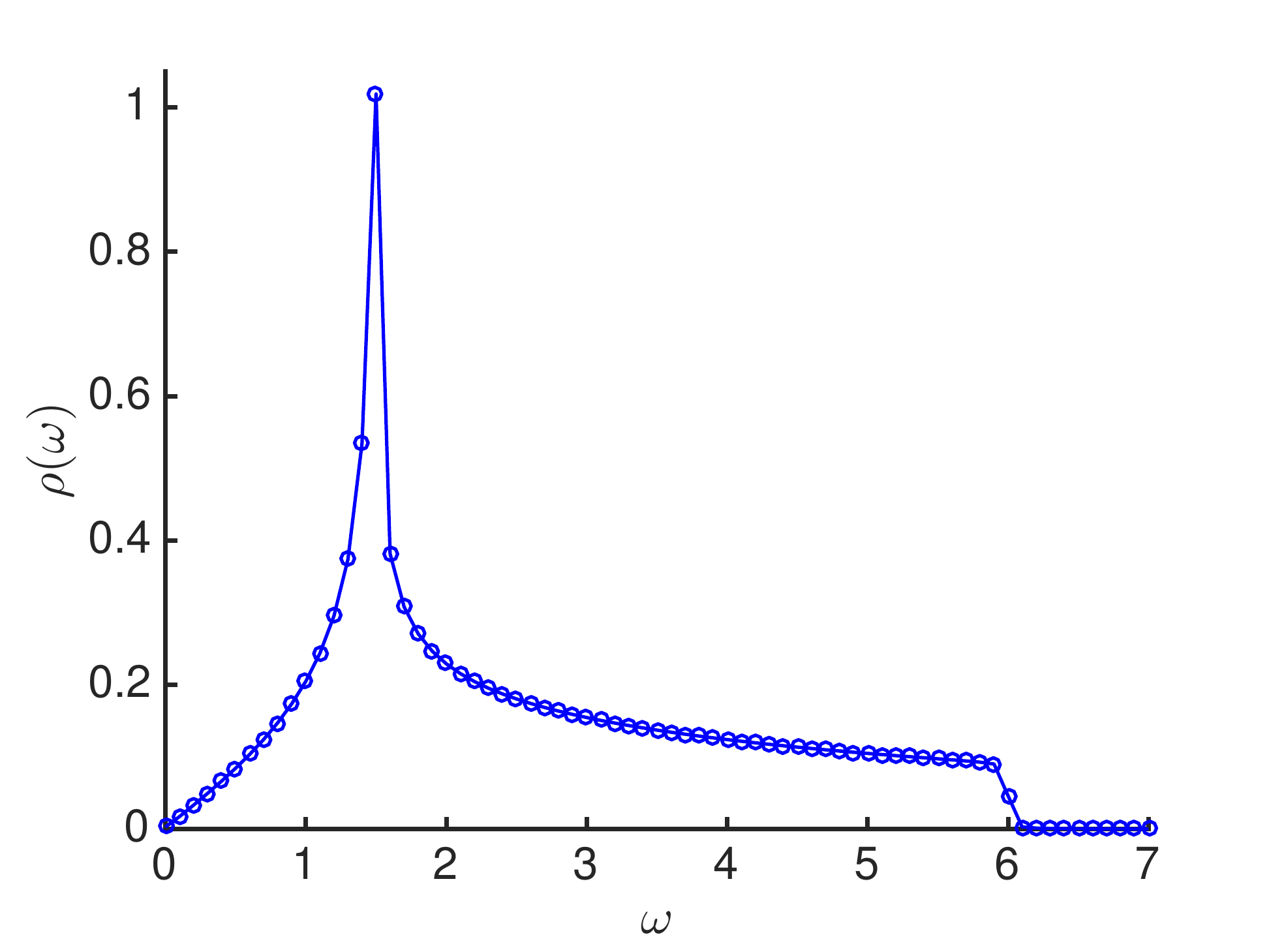}
\caption{Color online.  The density of states, $\rho(\omega)$, of the $Z_2$-invariant  pure-$K$ model. }
\label{dos}
\end{figure}
where $a=6K$ and $b=3K/4$. 
The maximum dispersion near $\bo=0$ is one of the distinctive features of this model as a result of pure $Z_2$ symmetry of the Hamiltonian.

  Figure \eqref{dos} shows the plot of the density of states for the pure-$K$ model, given by
\begin{align}
\rho(\omega)=\frac{1}{V}\sum_\bo \delta(\omega-\mathcal{E}(\bo)).
\end{align} 
The distinguishing feature of the density of states is that the largest peak correspond to  the saddle point of the excitation energy. The maximum excitation energy, however,  is at much higher energy and leads to a step-like van Hove singularity. The density of states also shows a discontinuity corresponding to the vanishing of the excitation spectrum at the midpoints of adjacent sides of the Brillouin zone in Fig.~\eqref{zone}. Thus, the $Z_2$-invariant pure-K model describes an unusual liquid. 
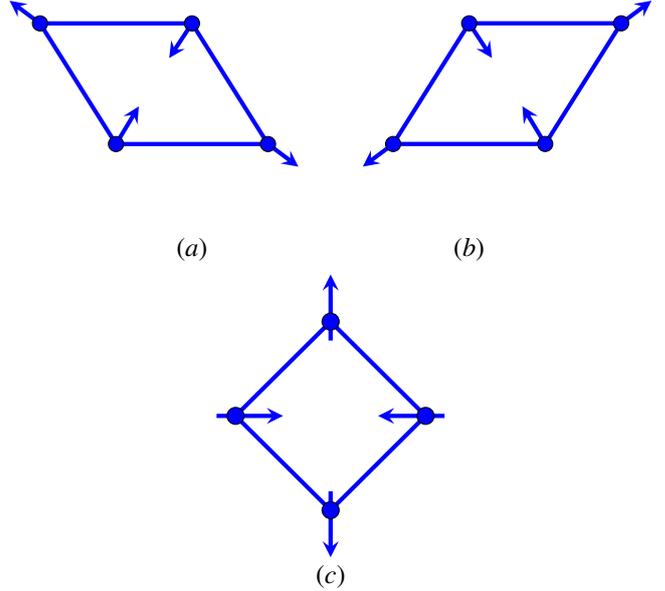
\begin{figure}[ht] 
\centering
\begin{tikzpicture}[scale=2] 
\draw[ultra thick,blue] (-1,0)--(-2,0) ;
\draw[ultra thick,blue] (-2.5,0.8)--(-1.5,0.8);
\draw[ultra thick,blue] (-2.5,0.8)--(-2,0);
\draw[ultra thick,blue] (-1.5,0.8)--(-1,0);
\draw [->,>=stealth,ultra thick, blue, blue]  (-1,0)--(-0.8,-0.15);
\draw [->,>=stealth,ultra thick, blue] (-2,0) --(-1.85,.25);
\draw [->,>=stealth,ultra thick, blue] (-1.5,0.8) --(-1.65,.57);
\draw [->,>=stealth,ultra thick, blue] (-2.5,0.8) --(-2.7,.95);
\draw[fill= blue] (-1,0) circle (0.5mm);
\draw[fill= blue] (-2,0) circle (0.5mm);
\draw[fill= blue] (-1.5,0.8) circle (0.5mm);
\draw[fill= blue] (-2.5,0.8) circle (0.5mm);
\draw(-1.5,-0.7) node[]{$(a)$};
\end{tikzpicture}
\quad \quad
\begin{tikzpicture}[scale=2] 
\draw[ultra thick,blue] (1,0)--(2,0) ;
\draw[ultra thick,blue] (2.5,0.8)--(1.5,0.8);
\draw[ultra thick,blue] (2.5,0.8)--(2,0);
\draw[ultra thick,blue] (1.5,0.8)--(1,0);
\draw [->,>=stealth,ultra thick, blue]  (1,0)--(0.8,-0.15);
\draw [->,>=stealth,ultra thick, blue] (2,0) --(1.85,.25);
\draw [->,>=stealth,ultra thick, blue] (1.5,0.8) --(1.65,.57);
\draw [->,>=stealth,ultra thick, blue] (2.5,0.8) --(2.7,.95);
\draw[fill= blue] (1,0) circle (0.5mm);
\draw[fill= blue] (2,0) circle (0.5mm);
\draw[fill= blue] (1.5,0.8) circle (0.5mm);
\draw[fill= blue] (2.5,0.8) circle (0.5mm);
\draw(1.5,-0.7) node[]{$(b)$};
\end{tikzpicture}
\quad \quad
\begin{tikzpicture}[scale=1.25] 
 \draw[ultra thick,blue] (1,0)--(0,1) ;
 \draw[ultra thick,blue] (0,1)--(-1,0);
 \draw[ultra thick,blue] (-1,0)--(0,-1);
 \draw[ultra thick,blue] (0,-1)--(1,0);
\draw [->,>=stealth,ultra thick, blue] (1.2,0) --(.5,0);
\draw [->,>=stealth,ultra thick, blue] (-1.2,0) --(-0.5,0);
\draw [->,>=stealth,ultra thick, blue] (0,.8) --(0,1.5);
\draw [->,>=stealth,ultra thick, blue] (0,-.8) --(0,-1.5);
\draw[fill= blue] (1,0) circle (0.9mm);
\draw[fill= blue] (-1,0) circle (0.9mm);
\draw[fill= blue] (0,1) circle (0.9mm);
\draw[fill= blue] (0,-1) circle (0.9mm);
\draw(0,-1.7) node[]{$(c)$};
\end{tikzpicture}
\caption{Color online.  The spin  configurations on the triangular plaquettes, which obey  the ``ice rules'' with two spins pointing inward and two spins pointing outward at each vertex. }
\label{ice}
\end{figure}

\section{Effects of frustration}
 We now consider the full model in Eq.~\eqref{model}. In the regime $J_z\ll K$, the physics of this Hamiltonian is, in fact,  the same as in the previous section. In the dominant Ising coupling, $J_z >K$, the sign of $K$ is very crucial and the system is frustrated as it is impossible to align the spins antiferromagnetically on the vertices of the triangular lattice. This leads to many classical degenerate ground states.  The classical degenerate ground states of the pure-$J_z$ term are known to be lifted  through order-by-disorder mechanism \cite{sow3} by quantum fluctuations emanating from the pure XY easy-axis ferromagnetic coupling $H_0 = -J\sum_{\la ij\ra}\lb S_{i}^+S_{j}^+ + S_{i}^-S_{j}^-\rb$.  In this case, quantum fluctuations select a particular state  known as a {\it ferrosolid} state \cite{sow3}. This state differs from the conventional {\it supersolid} state, \cite{G} as it breaks translational and $Z_2$ symmetries.  
 
 We can imagine covering the triangular lattice with plaquettes, then one of the  degenerate classical ground states of Eq.~\eqref{model} obeys the ice-rules depicted in Fig.~\eqref{ice}, in which the Ising term represents the degenerate classical ice and the ring exchange term denotes quantum fluctuations. However, the lifting of the classical  degeneracy by the ring exchange term  is a highly nontrivial mathematical problem. In fact, it is infeasible to analyze this problem analytically.   For $J_z>K$ and $K<0$, there is a possibility of a gapped QSL state  with gapped excitations on frustrated non-bipartite lattices.  Although we cannot  analytically confirm this claim, numerical techniques such as QMC should be tractable with $K<0$. It would be interesting to investigate numerically if Eq.~\eqref{model} promotes  a two-dimensional QSL state within a class of triangular lattice QSL materials, \cite{qsl,qsl1,qsl2}  and other QSL materials on the kagome \cite{balent} and pyrochlore lattices \cite{Huang, you,gin, zhi, sun1,sun4,sun4a, sun5,sun6,sun7,juan}.   A related U(1)-invariant  easy-axis model on the kagome lattice has been conjectured to possess a QSL phase \cite{bal}.

\section{Conclusion}
 In this communication, we presented the distinctive features of a $Z_2$-invariant XY ring exchange interaction on the triangular lattice.  We showed that the complete breaking of continuous U(1) symmetry down to discrete $Z_2$ symmetry has profound effects on the nature of the quantum properties that emerge from this system.  For the pure ring-exchange model with  $Z_2$-invariance, we showed that the distinguishing factor is the  gapped $\bo=0$ mode and soft modes at the midpoints of each side of the Brillouin zone.  As a result, the $Z_2$-invariant ring exchange model possesses some special features which are different from its U(1)-invariant counterpart. We also provided a glimpse into  the nature of the possible quantum phase that could emerge when competing with a classically frustrated Ising interaction. An explicit numerical simulation is required to uncover the nature of the proposed Hamiltonian and the possibility of any two-dimensional QSL states within a class of triangular lattice QSL materials such as  $\kappa$-(BEDT-TTF)$_2$Cu$_2$(CN)$_3$ and EtMe$_3$Sb[Pd(dmit)$_2$]$_2$  \cite{qsl,qsl1,qsl2, balent}. 
 
\section*{Acknowledgments} The author would like to thank African Institute for Mathematical Sciences (AIMS), where this work was conducted.  Research at Perimeter Institute is supported by the Government of Canada through Industry Canada and by the Province of Ontario through the Ministry of Research
and Innovation.

\end{document}